\documentclass[pra,aps,a4paper,twocolumn,10pt]{revtex4-1}
\usepackage{graphicx,amsmath,amssymb,datetime,braket}
\usepackage[usenames,dvipsnames]{color}

\newcommand{\figref}[1]{\text{Fig. \ref{#1}}}
\newcommand{\eqnref}[1]{\text{Eqn. \ref{#1}}}
\newcommand{\micron}{$\mu$m}

\begin{document}

\title{Prospects for using integrated atom-photon junctions for quantum information processing}
\author{R. A. Nyman}
\author{S. Scheel}
\author{E. A. Hinds}
\affiliation{Centre for Cold Matter, Blackett Laboratory, Imperial College
London, Prince Consort Road, SW7 2BW, United Kingdom}
\date{\today}

\begin{abstract}
We investigate the use of integrated, microfabricated photonic-atomic junctions
for quantum information processing applications. The coupling between atoms and light is enhanced by using microscopic optics without the need for cavity enhancement. Qubits that are collectively
encoded in hyperfine states of small ensembles of optically trapped atoms,
coupled via the Rydberg blockade mechanism, seem a particularly promising
implementation. Fast and high-fidelity gate operations, efficient readout, long
coherence times and large numbers of qubits are all possible.\end{abstract}

\maketitle

\section{Introduction}

Microfabricated structures for trapping and manipulating cold atoms, commonly
referred to as atom chips, offer obvious benefits for potential implementations
of quantum information processing (QIP) \cite{ReichelVuletic11}.
Microfabrication provides a straightforwardly scalable route to many qubits.
It is also possible to integrate more than one type of quantum-information
carrier such as superconducting microwave resonators, micro-mechanical
oscillators etc. onto a chip (see other articles in this special issue).

In this article, we examine why and how microfabricated optical elements are
useful, and how they may be integrated with atom chips for quantum information
processing purposes. We have recently demonstrated detection of cold atoms using
multiple channels of a monolithic optical waveguide chip \cite{Kohnen11a}, and
we will take these already-proven structures as a basis for application to QIP.

\begin{figure}
\includegraphics[width=\columnwidth]{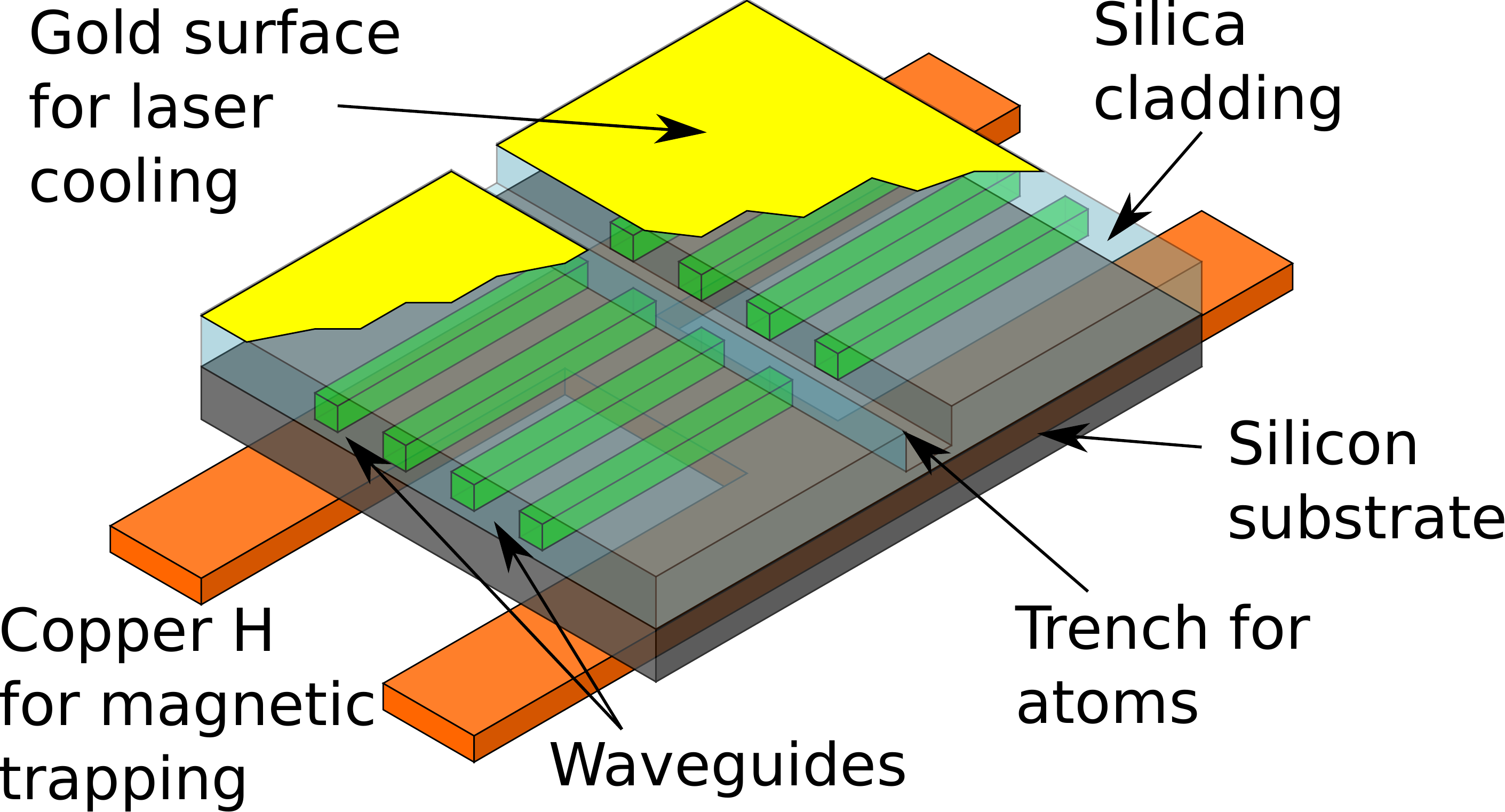}
\caption{Current chip design (not to scale). The integrated optical chip
consists of 12 buried waveguides in silica on a silicon substrate. Light is
brought to and from the chip by optical fibres (not shown). A trench cut into
the silica completes the atom-photon junctions, as described in
Ref.~\cite{Kohnen11a}. The whole chip is glued onto a macroscopic copper
structure, which carries the current for magnetic and magneto-optical
trapping.}
\label{Fig: chip schema}
\end{figure}

The device we consider is a microfabricated optical chip integrated together
with a current-carrying chip for magnetic trapping and evaporative cooling (see
\figref{Fig: chip schema}). For the purposes of this article, we take as the
starting point that an ultracold atom cloud (potentially a Bose-Einstein
condensate) is held in the trench of a photonic chip for a few seconds, and
aligned with respect to the photonic chip, to form an array of well coupled
atom-photon junctions. The magnetically-trapped sample of atoms can then be
manipulated optically for all quantum information processing operations.

Throughout this article, we will assume that the qubits are to be encoded in
the hyperfine states of an atomic sample, in particular the $\ket{F=2}$ and
$\ket{F=1}$ states of the $^{87}$Rb $5S_{1/2}$ ground state. To form one qubit, a sample of up to a few hundred atoms at a temperature of a few microkelvin will be trapped inside the trench of the atom chip by light from one waveguide. We will use a collective encoding in which we take the logical qubit state $\ket{0}$ to be the one in which each atom is prepared in its $\ket{F=1}$ hyperfine ground state,
\begin{equation}
\ket{0} = \ket{F_1=1,F_2=1,F_3=1,\ldots}\,,
\label{eq:logical0}
\end{equation}
where $F_i$ labels the hyperfine state of the $i$\textsuperscript{th} atom.
In a collective encoding, the logical qubit state $\ket{1}$ will contain a
single excitation of $\ket{F=2}$ that is shared among all atoms,
\begin{gather}
\label{eq:logical1}
\ket{1} =\frac{1}{\sqrt{N}}
\Big[
\ket{F_1=2,F_2=1,F_3=1,\ldots}\hspace{7ex}\\
\hspace{15ex}+\ket{F_1=1,F_2=2,F_3=1,\ldots} +
\cdots \nonumber
\Big] .
\end{gather}

Multiple qubits will be individually addressed using light carried by waveguides of a microfabricated photonic waveguide chip, as illustrated in Fig.~\ref{Fig: chip schema} and described in Ref.~\cite{Kohnen11a}. Qubit manipulations and readout are
to be purely optical. Specifically, we will investigate the use of the Rydberg
blockade mechanism for both one- and two-qubit operations.

This article is organised as follows. In Sec.~\ref{sec:setup} we describe our present chip design and briefly review the fundamental limits to the measurement of atomic density. We then discuss in Sec.~\ref{sec:implementation} the implementation of QIP on our chip, focussing on the requirements for qubit initialisation, manipulation and detection. As the present chip design is preliminary, we suggest modifications to the chip in Sec.~\ref{sec:modifications} that may lead to better performance.

\section{Experimental setup}
\label{sec:setup}

\subsection{Present chip design}

Our current array of integrated atom-photon junctions, as described in
Ref.~\cite{Kohnen11a}, includes 12 optical waveguides on a 10~\micron\ pitch at
the centre of the chip. They are cut across by a 16~\micron\ wide trench for
atom access, and the waveguides are 10~\micron\ below the gold-coated chip
surface. The waveguides are buried and are formed by doped silica in silica on a
silicon substrate. The doping gives a 0.75\% refractive index step, and the
4~\micron\ square waveguides support a single transverse mode (with a mode-field
$1/e$ radius of 2.2~\micron) at 780~nm, resonant with the rubidium $D_2$
transitions. The waveguides fan out to a 250~\micron\ pitch at the edge of the
chip, where they are connected using commercial arrays of fibres, mounted in V-groove
structures. We use optical-index-matching glue Epotek OG-116 for the
connections.

We have measured the absorption of resonant light carried by the
waveguides due to the presence of a low density ($\sim 10^{-2}$ atoms per
\micron$^3$) of laser-cooled $^{87}$Rb atoms (at around 100~$\mu$K) being
launched through the trench. We have also used the atoms to probe the intensity
and polarisation of light carried by the waveguides. As a result, we know that
the chip has good polarisation maintaining properties, and can easily be
calibrated for atom number measurements.

The magnetic fields for magnetic and magneto-optical traps are generated using a
current-carrying sub-chip. The various layers of the chip are machined from
copper sheet, and separated and held in place using machinable aluminium
nitride and Epotek H-74 adhesive. The materials are chosen to maximise thermal
conduction, and we routinely use currents up to 80~A, thereby trapping up to
20~million atoms. Careful choice of adhesives is essential to ensure good enough
vacuum for efficient evaporative cooling, along with good optical connections.

The current chip is not ideal for QIP but it provides a useful starting point for considering the subject. In future chips we may adjust the waveguide mode size, trench width, and magnetic trap design. The number of channels is
scalable well beyond the current twelve, but that is already enough for demonstrations of QIP. It is also
entirely practical to include other components as well as waveguides on the chip, e.g. switches, beamsplitters and optical interferometers, as we discuss briefly in Sec.~\ref{sec:modifications}.

\subsection{Advantages of microfabricated optics: fundamental limit to the
measurement of atomic density}

Readout of an atom-encoded qubit typically requires state-sensitive atom detection,
with a minimum of lost qubits. However, detection by passing light through an atomic sample improves when more photons are scattered.

Consider scattering light at the atomic resonance frequency $\omega$, with
a local atom number density $\rho(x,y,z)$. The light power scattered by the atom
cloud (for intensity far below saturation intensity, and very small optical
thickness) is
\begin{equation}
P_{sc}=\frac{\hbar\omega\Gamma}{2I_{sat}}\int\,dx\,dy\,dz\,I(x,y)\rho(x,y,z)\,\label{eqn:Psc}
\end{equation}
Here the beam propagates along the $z$-direction, $I$ is its intensity; and
$\Gamma$, $I_{sat}$ and $\hbar \omega$, respectively, are the natural full width
at half maximum, saturation intensity and energy of the atomic transition. The prefactor in \eqnref{eqn:Psc} is the optical scattering cross-section,
\begin{eqnarray}
\sigma = \frac{\hbar \omega \Gamma}{2 I_{sat}}.
\end{eqnarray}

We define the effective beam area $A$ and the effective number of atoms in the beam $N_{at}$ by taking intensity-weighted averages, normalised to the peak intensity $I_0$,
\begin{eqnarray}
A &=& \frac{1}{I_0}\int\,dx\,dy\,I(x,y)\,,\\
N_{at} &=& \frac{1}{I_0}\int\,dx\,dy\,dz\,I(x,y)\rho(x,y,z)\,.
\end{eqnarray}
For a Gaussian beam of mode field radius $w$, the effective area is
$A=\pi w^2 / 2$. If the atom cloud is much broader than the beam, such that we
can set $\rho(x,y,z)=\rho(z)$, then $N_{at} = A\int dz\, \rho(z)$.

In a transmission experiment with $N_\gamma$ incident photons, the number of transmitted photons is
$N_{det} = N_\gamma\left(1-\frac{\sigma}{A}\,N_{at}\right)$ (we assume $\frac{\sigma}{A}{N_{at}}\ll 1$).
The noise in this number is $\sqrt{N_{det}} \simeq \sqrt{N_\gamma}$ (with weak absorption), due to the Poissonian
statistics of photon arrival. Hence the uncertainty in the atom number derived from this signal is $\sigma_{N_{at}} = A/(\sigma\sqrt{N_\gamma})$. In terms of the number of scattered photons per atom, $n_{sc}=(\sigma/A)N_{\gamma}$, this becomes
\begin{eqnarray}\label{Eq:sigmaN}
\sigma_{N_{at}} = \sqrt{\frac{A}{\sigma n_{sc}}} \,.
\label{Eqn: atom number uncertainty}
\end{eqnarray}

The same uncertainty in the measured number of atoms can be derived
for optical phase shift measurements using a balanced Mach-Zehnder setup (two photo-detectors, counting nearly the same number of photons).  We conclude that the
uncertainty in the measured number of atoms depends on the destructiveness, characterised by $n_{sc}$  (photon scattering tends to destroy qubit coherence as well as leading to atom loss) and on the ratio of beam area to scattering
cross-section \cite{Horak03,Lye04}. For a given value of $n_{sc}$, small beams give a more accurate readout of the atom number and therefore give better state-selective qubit readout. This is a key advantage of microfabricated optics. Our optical waveguides are well suited to the recently-demonstrated method of two-frequency interferometric detection, which is minimally destructive\cite{Lodewyck09,Bernon11, Kohnen11b}.

\section{A possible implementation of QIP via the Rydberg blockade}
\label{sec:implementation}

\begin{figure}
\centering
\includegraphics[height=11cm]{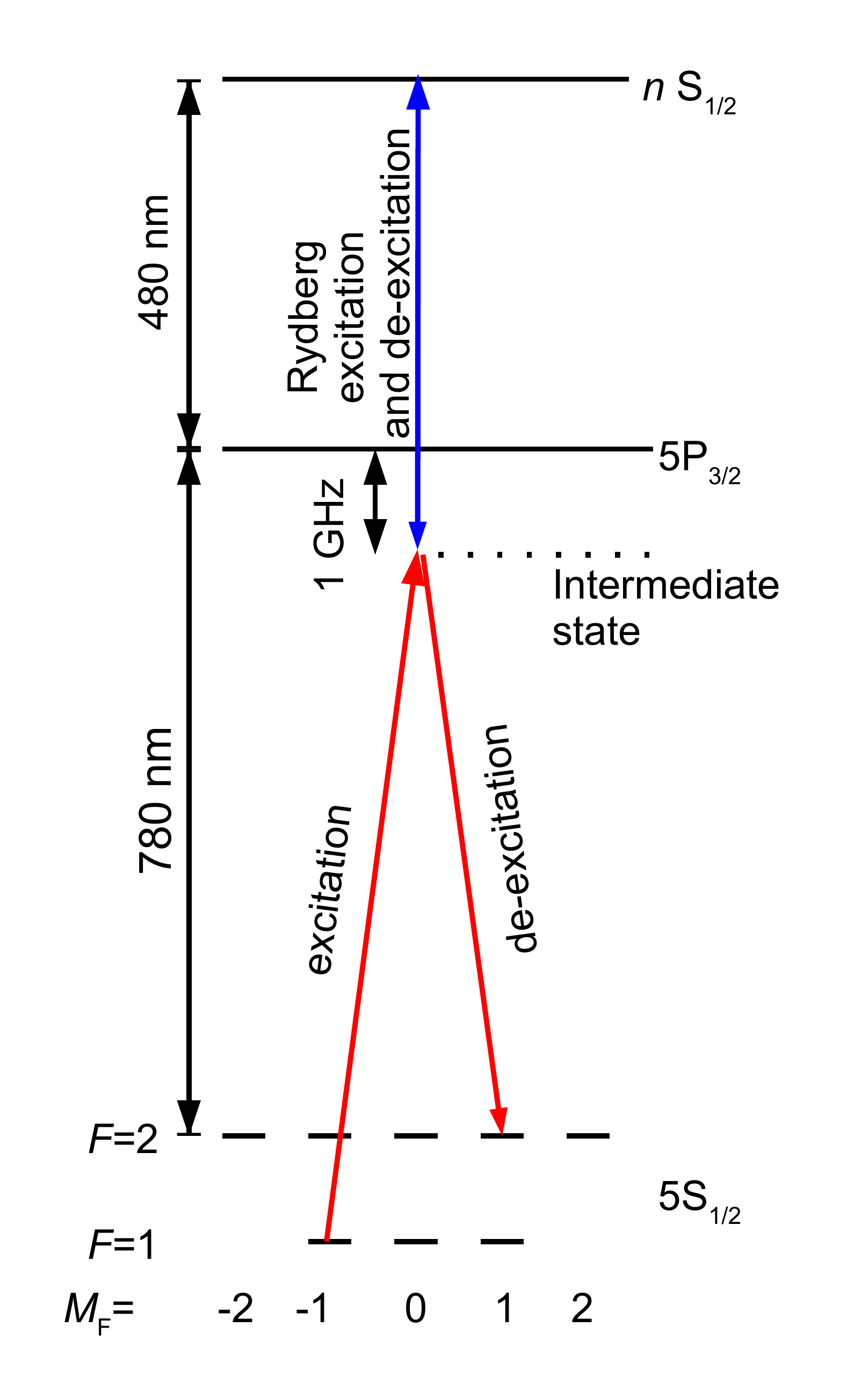}
\caption{Example gate operation: the Hadamard gate rotates from a state having all
atoms in $\ket{F=1,m=-1}$ to a shared excitation of one atom in $\ket{F=2,m=+1}$. The
single excitation is guaranteed by the Rydberg blockade mechanism. The 480~nm light
comes from outside the chip, while the 780~nm light is carried by optical waveguides and
provides site selection. Principal quantum numbers of $n\simeq 40$ can be used
for one-qubit operations.}
\label{Fig: excitation}
\end{figure}

Collectively-encoded qubit operations can be achieved using the Rydberg blockade
mechanism \cite{Lukin01}, which permits the strength of interactions between
atoms to be switched over up to 12 orders of magnitude \cite{Saffman10}. The
protocol starts by loading atoms from a magnetic trap into optical traps that
are generated and addressed by individual waveguides. Two-photon excitation to Rydberg states will use light of 780~nm wavelength carried by the waveguides, together with 480~nm light from outside the chip, covering all the atom-photon junctions, as shown in \figref{Fig: excitation}. A universal set of gate operations requires only one kind of two-qubit operation, but two one-qubit operations. The Rydberg blockade mechanism provides the switchable interactions needed both for the
two-qubit gate, and one of the one-qubit rotations. As for any proposed QIP implementation, we ask the question: is it good enough to perform real calculations?

In the following, we discuss in detail how our proposed system would fulfill standard criteria \cite{DiVincenzo00} for implementing QIP. The criterion of scalability is satisfied up to a point by the multi-channel nature of the
microfabricated optical chip, which allows interaction between nearest neighbours on a line.

\subsection{Microscopic traps (qubit initialisation)}

It must be possible to initialise the qubits to a simple known state, typically with all qubits in logical $\ket{0}$. Initialisation of a qubit in our system means loading a well-defined number of atoms from the magnetic trap into an optical trap formed by off-resonant light carried by a waveguide channel. The channels are two-sided, and maintain polarisation \cite{Kohnen11a}, making it possible to produce an optical lattice in the trench, or an interference-free trap from the two beams, using a slight frequency difference. Optical traps are not loaded directly from the magneto-optical trap, as the atomic density is too low.

Starting from a thermal cloud of around $10^5$ atoms at 2~$\mu$K in a magnetic
trap with axial (radial) frequency 20~Hz (1~kHz), the cloud is 220~\micron\ long
($1/e^2$ density), with a central linear density of 360 atoms per micron, and
transverse size 2.2~\micron\ (well matched to the mode field size of the optical
waveguide). Under these conditions, up to 1500~atoms could be loaded into a
non-lattice trap using $2\times 80~\mu$W of trapping light at 830~nm, scattering
no more than 1 photon per second, and with acceptable collisional 3-body loss
rates.

In order to achieve the best mode-matching between the optically trapped cloud
and the waveguide mode for detection, radial and axial trapping frequencies can
be chosen independently by choosing trapping power and the degree of interference between the two
counter-propagating beams. In the example above, without interference the
axial and radial trapping frequencies are 0.3\,kHz and 6.6~kHz; with full
interference they are 120~kHz and 9~kHz, respectively.

\subsection{Readout efficiency with a waveguide chip}\label{Sec: readout}

At some point in the processing of quantum information, it becomes necessary to have qubit-specific state readout. For qubits encoded in the hyperfine ground states of atoms, this translates into state-selective, localised, atom detection. When detection is through the absorption of a single probe laser beam the number of scattered photons per atom is typically limited to roughly 100 \cite{Bochmann10}. Given the mode size of the present waveguide chip, and the over-all transmission and detection quantum efficiency of 20\%, and restricting the average number of scattered photons to 100 per atom, readout of a single atom can be performed with a signal-to-noise ratio of 1.

If the light is reflected back and forth across the trench by mirrors of power reflectivity $R$, for example by coating the ends of the waveguides, the optical cross section is effectively increased by a factor of $1/(1-R)$. This reduces the uncertainty in the measured atom number by a further factor of $\sqrt{1-R}$ compared with \eqnref{Eq:sigmaN}. Thus a reflectivity of $0.9$ would be sufficient to improve the signal-to-noise ratio to 3. However, the cavity formed by plane parallel mirrors is unstable. The light spreads through diffraction, giving an effective reflectivity that is of order the Rayleigh length ($\pi w_0^2/\lambda\simeq20\,\mu$m) divided by the width of the trench (16\,\micron\ in our case). The plane Fabry-Perot geometry does not offer any significant improvement in sensitivity unless the trench is made much narrower than the Rayleigh length, and then it is too narrow to contain the Rydberg atoms \footnote{Gleyzes \textit{et al.} \cite{Gleyzes09} have produced gapped cavities with satisfactory finesse, despite using a more lossy material than silica, since their trench was just 2~\micron\ across.}. It is not known at present how to form concave ends on the integrated waveguides, but this could form a stable cavity, which would be very helpful for improving the detection performance. Taking all these points together, it does not seem possible to to achieve pure on-chip readout without an upgrade of our present chip. We therefore consider an alternative method of readout.

Our alternative is to measure fluorescence using a camera outside the trench. Light, red-detuned by 6~MHz from the cycling transition (ground state $\ket{F=2}$ to excited $\ket{F'=3}$) can be sent into both sides of one channel, to laser cool while the camera detects the scattered photons. With this two-sided method many more than photons can be scattered than the 100 of the single-probe transmission experiment. A conservative numerical aperture of 0.33 (i.e. a 35~mm diameter lens 100~mm from the chip, probably outside the vacuum chamber) will collect 1\% of the light. Single sites will be easily resolvable, and cameras with quantum efficiencies in excess of 50\% are available. On average, 6000 spontaneous emission events can occur over 360\,$\mu$s before unwanted state depumping occurs, by which time the camera detects 30 counts. Therefore, the readout can achieve very high fidelity, as required. In addition, the total atom number used for each qubit can be accurately measured.

\subsection{Single-qubit rotations}

The two, one-qubit operations we choose for our universal set of gates are the
phase-shift gate
\begin{equation}
\label{eq:psg}
a\ket{0}+b\ket{1}\,\rightarrow\,a\ket{0}+b{\rm e}^{{\rm i}\phi}\ket{1}
\end{equation}
and the Hadamard gate
\begin{equation}
\label{eq:hadamard}
\ket{0}\,\rightarrow\, \frac{1}{\sqrt{2}}\left(\ket{0}+\ket{1}\right) \,.
\end{equation}
The phase-shift gate (\ref{eq:psg}) requires a state-dependent potential,
which can be supplied by a laser whose wavelength is significantly closer to
resonance with one hyperfine ground state than the other. For example, laser
light of 250~nW power detuned by 20~GHz from the
$\ket{F=2}\rightarrow\ket{F'=3}$ transition of the $D_2$ line (780~nm), sent
down a waveguide, would give a relative light shift of about $h\times 0.24$~MHz
($h$ being Planck's constant) between the two hyperfine ground state levels, and
would lead to $\pi/2$ phase gate times of $1~\mu$s. Only 0.0015 photons per gate
operation would be scattered per atom.

\figref{Fig: excitation} illustrates the basic scheme used to implement the second type of single-qubit rotation, the Hadamard gate of \eqnref{eq:hadamard}. The logical $\ket{0}$ state (\eqnref{eq:logical0}) is coupled to logical $\ket{1}$ (\eqnref{eq:logical1}) by a 4-photon interaction, which is applied for long enough to constitute a $\pi/2$ rotation on the Bloch sphere describing this logical state space. The first two photons, one at 780\,nm and the other at 480\,nm, resonantly excite one of the atoms in the $\ket{0}$ ensemble to a high-lying Rydberg state $nS_{1/2}$. The 2-photon excitation rate is enhanced by virtue of the intermediate $5P_{3/2}$ level, but this level is sufficiently far-detuned that no significant population appears in the $5P_{3/2}$ state. The repulsive interaction between the Rydberg $nS$ states \cite{Heidemann07} ensures that only one atom can be resonantly excited. The third and fourth photons resonantly couple this Rydberg excitation back down to the logical $\ket{1}$ state. We note that the trapping light should be switched off during gate operations, to avoid photo-ionisation. The circularly-polarised red (780~nm) light is carried by the optical waveguides, while the linearly-polarised blue (480~nm) light is shone in from outside the trench. With asymmetric focussing down to
waists of 10~\micron\ and 80~\micron, respectively, multiple sites can be
addressed at will; the waveguide-transmitted red light provides the necessary
site selectivity.

Using up to 80~mW at 480~nm, and 25~nW of 780~nm (1~GHz intermediate-state detuning), the collectively-enhanced Rabi frequency for excitation to the $n\simeq40$ Rydberg state can be up to $\sqrt{N}\times500$~kHz (depending on polarisations). Assuming as many as $N=500$ atoms encoding the qubit, the Hadamard operation, can be as short as 25~ns, which is much faster than the excited state natural lifetime (around $100~\mu$s for a single atom with $n\simeq40$). Methods of control have been developed by the NMR community to achieve accurate $\pi/2$-pulses despite the uncertainty in the atom number, and hence in the collective Rabi frequency \cite{Vandersypen05}.

The $n\simeq40$ Rydberg-state energy shift due to the presence of an $n\simeq40$
Rydberg-state atom is at least 90~MHz for a qubit with spatial extent of
2~\micron\,\footnote{Temporary axial compression of the qubit may be required to
achieve that size, performed by increasing the trapping light intensity. }.
Because this shift is much larger than the power-broadened linewidth
(i.e. greater than both the excited-state natural linewidth and the Rabi
frequency for excitation), the excitation of a second atom to the Rydberg state
is suppressed by being off-resonant.

\subsection{Two-qubit gate}

We propose to implement a controlled phase gate as the required two-qubit operation, using the Rydberg blockade mechanism once again. On the present atom chip, neighbouring qubits are separated by 10\,\micron\ and therefore the operation requires a larger blockade radius than the 2\,\micron\ of the Hadamard gate discussed above. This can be achieved using a higher principle quantum number, $n\simeq100$, compared with $n\simeq40$ for the Hadamard gate. The Rydberg blockade shift for two trapped single-atom qubits has already been
used to produce phase gates and entanglement \cite{Urban09,Isenhower10,Wilk10}.
The blockade range reported there was up to 10~\micron, exactly the
inter-qubit spacing of our present chip, giving an energy shift of over 50~MHz in one
qubit due to the other.

The excitation once again uses a combination of red and blue light, this time tuned to drive state $\ket{1}$ up to $n=100$ and back. When the control qubit is in state $\ket{1}$, the light excites the Rydberg state and this blocks any excitation of the target qubit, which is therefore left unchanged at the end of the operation. By contrast, when the    control qubit is in state $\ket{0}$, the light does not excite it. The target in state $\ket{0}$ also not excited, but state $\ket{1}$ undergoes a full $\pi$ rotation that converts it to $-\ket{1}$, as required for a phase gate.

An additional feature that has to be taken into account in our setup is the
interaction of the trapped atoms with the dielectric trench wall that will be
only 8~\micron\ away from the centre of the atomic sample. Measurements
\cite{Sandoghdar92,Kuebler10,Tauschinsky10} as well as theoretical studies \cite{Hinds91,Crosse10}
have shown that the dispersion interaction with a wall leads to energy shifts
in Rydberg atoms that can exceed several MHz depending on the wall material
(in Ref.~\cite{Kuebler10} shifts of the $43S_{1/2}$ state as high as 200~MHz
were reported inside a quartz wedge cell of similar extent filled with hot Rb
vapour). The measured line shifts and widths depend sensitively on the chosen
Rydberg state and the surface plasmon structure of the dielectric wall. At
present it remains an open problem to determine the largest line width (or equivalently the
highest principal quantum number) that can be used for QIP.

\subsection{Ultimate performance and fidelity limits}

Qubit initialisation, imperfections of gate operation and readout, and qubit
decoherence during computation all influence the fidelity of a quantum
information processing device. Saffman and Walker \cite{Saffman05} have analysed in
detail the sources of gate infidelity and loss of coherence in a 2-qubit
Rydberg-atom quantum computer. We follow their logic to estimate ultimate limits
to fidelity, assuming that technical limits can be overcome.

A long coherence time depends on choosing a suitable pair of magnetic sublevels in which to encode the qubits. We suggest using $\ket{F=1,m=-1}$ and $\ket{F=2,m=1}$ with a magnetic field of 3.23\,G directed across the trench, in the direction of the waveguides. It has been shown \cite{treutlein04} that superpositions of these states in the presence of this field can remain coherent for several seconds.  The gold mirror surface of the optical chip, some 10\,\micron\ away from the atoms, induces spin flips \cite{Rekdal04} at approximately 0.5~s$^{-1}$ in a field of a few Gauss. Spontaneous scattering of the  830~nm trap light decoheres the qubit over a time of order 1\,s.  Saffman and Walker~\cite{Saffman05} show that decoherence due to inhomogeneous ac Stark shifts is slower than 1\,s$^{-1}$. These are the main anticipated causes of decoherence. Thus the gate operation is 5--6 orders of magnitude faster than the decoherence time of the qubit.

If the excitation time is optimally chosen, the minimum, averaged Rydberg gate error is approximately $3(B\tau)^{-2/3}$ where $B$ is the blockade (angular) frequency shift and $\tau$ is the radiative lifetime of the Rydberg state \cite{Saffman10}. For a Rydberg state with $n\simeq100$, an excitation time of 10~$\mu$s gives a minimum error of about 1\%. This is due to a combination of spontaneous emission from the Rydberg state, and population left in the Rydberg state at the end of the gate. Spontaneous emission causes errors of around $10^{-3}$ in the phase gate. The computational qubit space covers only the $\ket{F=1,m=-1}$ and $\ket{F=2,m=1}$ states. If the circular polarisations of the two red photons are imperfect, then qubits are effectively lost. However, experiments on our photonic chip have already indicated that adequate purity of polarisation can be achieved \cite{Kohnen11a} to reach fidelities better than 99\%. To summarise, it seems realistic that the fidelity can be about 99\%, principally limited by the Rydberg state lifetime.

\section{Modifications to the chip}
\label{sec:modifications}

\subsection{Optical components: switches and interferometers}
\label{sec: optical components}

\begin{figure}
\includegraphics[width=\columnwidth]{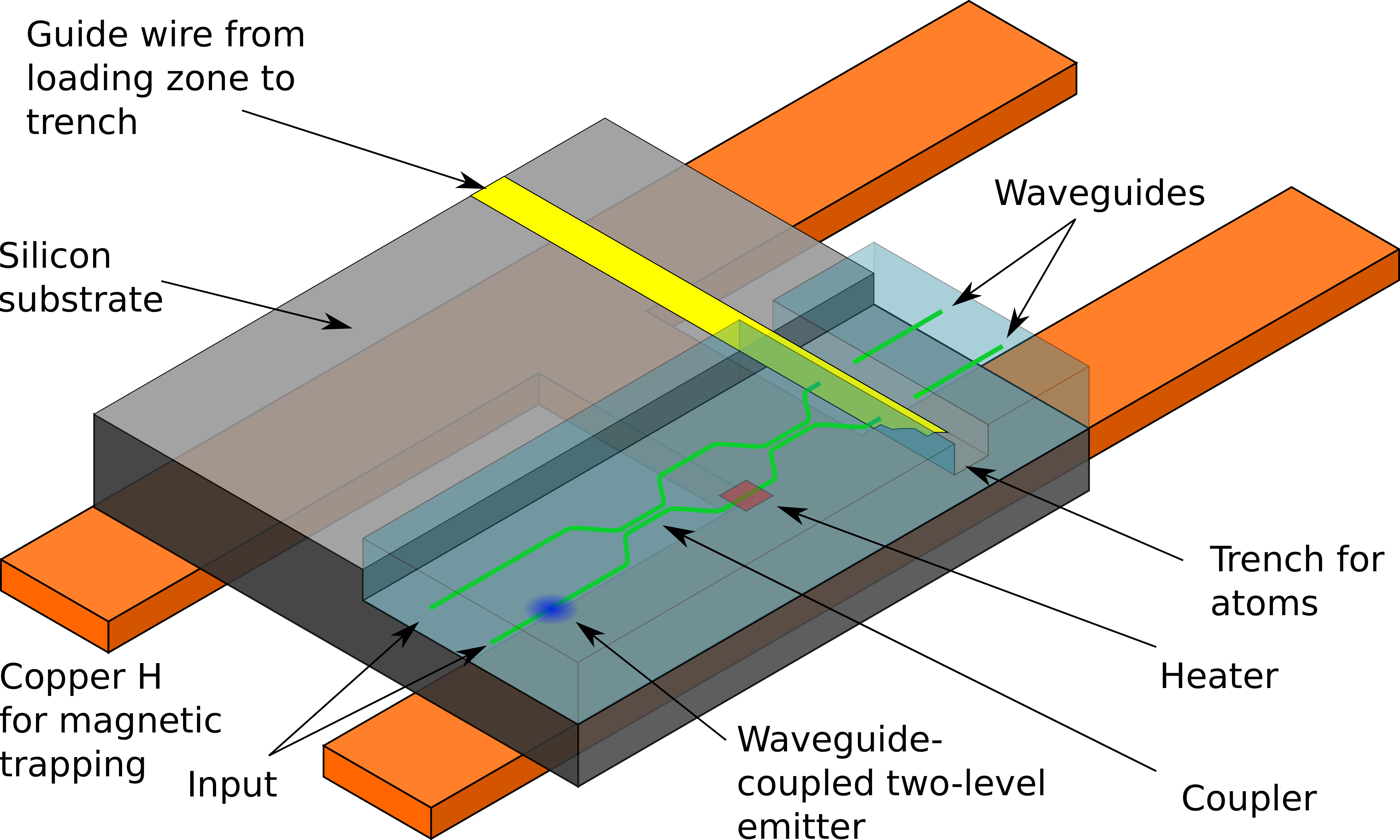}
\caption{Some components available for a future photonic-atom chip: A Mach-Zehnder interferometer, using directional couplers, with a heater in one arm, switches inputs and outputs for a many-in, many-out optical system. It may be possible to integrate two-level emitters (e.g. single molecules or quantum dots) which are strongly coupled to the optical waveguides. Atoms could be loaded in one zone and guided across the chip with integrated wires, to the atom-photon junctions.}
\label{Fig: future chip}
\end{figure}

It is possible to integrate other optical elements into a chip to produce a more functional device. \figref{Fig: future chip} shows an example of the kind of integration we have in mind, where atoms on the chip interact with a complex optical circuit. As in the current version, an array of fibres brings light to several parallel waveguide channels indicated in the figure. In one or more of these waveguides it is also possible to incorporate individual emitters of single photons as an internal resource for QIP on the chip. These integrated emitters could be trapped atoms or ions, but are more likely to be solid-state optical quantum systems such as quantum dots or single molecules, strongly coupled to the waveguides \cite{Hwang11}.

The light propagates through a coupling region, where evanescent tunnelling splits the photon mode between neighbouring waveguides. Two couplers in sequence form an interferometer, whose path difference can be controlled by a local heater (see for example Ref.~\cite{Kawachi90}) as shown in Fig.\,\ref{Fig: future chip}, allowing a photon to be delivered to either channel or to a superposition of both. Such interferometers have been used for all-optical quantum information processing at 804\,nm by the group of J.~O'Brien \cite{Politi08} among others and could readily be adapted to the Rb wavelengths.

Operation of the gates requires control over the
frequency and intensity of many light beams used for optical pumping, trapping, gate operations and detection. This is an engineering problem well suited to large-scale optical integration. A monolithic, on-chip switchyard should be able scale up the operation to many gates. The ultimate limit to the number
of channels will probably be set by the need to supply of ultracold atoms for
each qubit, rather than by the limit of optical scalability.

It would be helpful to reduce the distance between qubits in a future chip so that lower principle quantum numbers could be used in the two-qubit Rydberg blockade gates. In our present chip, the optical channels are separated by 10\,\micron\ in order to avoid unwanted cross-talk through evanescent coupling. In future we will be able to reduce this distance by using waveguide material with a larger refractive index (e.g. SiO$_x$N$_y$) to confine the mode more strongly. For example, with 1.5\% index step, the channel spacing can be reduced to 8~\micron\ with the same level of cross-talk as the present chip.

In our current setup, the optical chip is glued onto the subchip containing the current-carrying wires of the magnetic atom trap. The tolerance on the alignment is small enough that it would be preferable to trap the atoms using wires microfabricated directly into the photonic chip, as illustrated in Fig\,\ref{Fig: future chip}. For example, a wire located at the bottom of the trench ($\sim10\,$\micron\ from the atoms) could be $\sim15\,$\micron\ across, which would carry enough to carry current to trap the atoms. Such a current-carrying wire would allow accurate transfer from a larger magnetic
trap or could work as a conveyor \cite{Long05} to bring atoms from a
loading zone on the chip.\\

\subsection{Other qubit systems}

Another advantage of microfabricated atom-optical systems, in addition to scalability and miniaturisation, is their suitability for hybrid systems, where two or
more types of qubit are brought together (see other articles in this special
edition). If the connections to the photonic chip can stand being cooled to
cryogenic temperatures, a superconducting microwave circuit may be
strongly coupled to an atom cloud \cite{Petrosyan09}. The protocol proposed in
Ref.~\cite{Petrosyan09} uses Rydberg states of an atom cloud about 10~\micron\
from the wires.

Another possibility is to couple atoms to an integrated, high-finesse optical cavity (e.g. Ref \cite{Trupke07,Colombe07}) to take advantage of high-co-operativity cavity QED for improved readout or coupling to flying qubits \cite{Horak03}.

\section{Conclusions and outlook}
\label{sec:conclusions}

We have shown that integrated atomic-photonic chips can be useful for quantum
information processing. Collectively-encoded atomic hyperfine states provide a
promising basis for qubits, with high-lying Rydberg levels being used as
intermediate states for gate operations. Gate and readout fidelities are
potentially as large as 99\%, and qubit numbers can be large (12 or more). Gate operations
are 5-6 orders of magnitude faster than intrinsic decoherence processes.

Interactions between atoms in high-lying Rydberg states and the
trench walls are still not fully characterised. It is known that resonant coupling
to surface plasmons can increase line widths enormously \cite{Kuebler10}.
Hence, the choice of intermediate Rydberg state for the gate operations is strongly linked with the choice of chip material.

Improved fabrication procedures together with an optical switchyard will permit even more than the current 12 atom-photon junctions to be used for QIP. There is no reason why alternative qubit systems could not be used alongside neutral atoms, with the same optical technology.

\begin{acknowledgements} 
This work was supported by EPSRC (UK), the Royal Society (UK), and European Projects HIP and AQUTE.
\end{acknowledgements}


\begin{thebibliography}{10}

\bibitem{ReichelVuletic11}
{\em Atom Chips}, edited by J. Reichel and V. Vuletic (Wiley-VCH, {}, 2011).

\bibitem{Kohnen11a}
M. Kohnen, M. Succo, P.~G. Petrov, R.~A. Nyman, M. Trupke, and E. A.Hinds,
  Nature Photonics {\bf 5},  35  (2011).

\bibitem{Horak03}
P. Horak, B.~G. Klappauf, A. Haase, R. Folman, J. Schmiedmayer, P. Domokos, and
  E.~A. Hinds, Phys. Rev. A {\bf 67},  043806  (2003).

\bibitem{Lye04}
J.~E. Lye, J.~J. Hope, and J.~D. Close, Phys. Rev. A {\bf 69},  023601  (2004).

\bibitem{Lodewyck09}
J. Lodewyck, P.~G. Westergaard, and P. Lemonde, Phys. Rev. A {\bf 79},  061401
  (2009).

\bibitem{Bernon11}
S. {Bernon}, T. {Vanderbruggen}, R. {Kohlhaas}, A. {Bertoldi}, A. {Landragin},
  and P. {Bouyer}, ArXiv  1103.1722  (2011).

\bibitem{Kohnen11b}
M. Kohnen, P.~G. Petrov, R.~A. Nyman, and E.~A. Hinds, arXiv:1104.0236  (2011).

\bibitem{Lukin01}
M.~D. Lukin, M. Fleischhauer, R. Cote, L.~M. Duan, D. Jaksch, J.~I. Cirac, and
  P. Zoller, Phys. Rev. Lett. {\bf 87},  037901  (2001).

\bibitem{Saffman10}
M. Saffman, T.~G. Walker, and K. M\o{}lmer, Rev. Mod. Phys. {\bf 82},  2313
  (2010).

\bibitem{DiVincenzo00}
D.~P. DiVincenzo, Fortschritte der Physik {\bf 48},  771  (2000).

\bibitem{Bochmann10}
J. Bochmann, M. M\"ucke, C. Guhl, S. Ritter, G. Rempe, and D.~L. Moehring,
  Phys. Rev. Lett. {\bf 104},  203601  (2010).

\bibitem{Note1}
Gleyzes \protect \textit {et al.} \cite {Gleyzes09} have produced gapped
  cavities with satisfactory finesse, despite using a more lossy material than
  silica, since their trench was just 2~$\mu $m\ across.

\bibitem{Heidemann07}
R. Heidemann, U. Raitzsch, V. Bendkowsky, B. Butscher, R. L\"ow, L. Santos, and
  T. Pfau, Phys. Rev. Lett. {\bf 99},  163601  (2007).

\bibitem{Vandersypen05}
L.~M.~K. Vandersypen and I.~L. Chuang, Rev. Mod. Phys. {\bf 76},  1037  (2005).

\bibitem{Note2}
Temporary axial compression of the qubit may be required to achieve that size,
  performed by increasing the trapping light intensity.

\bibitem{Urban09}
E. Urban, T.~A. Johnson, T. Henage, L. Isenhower, D.~D. Yavuz, T.~G. Walker,
  and M. Saffman, Nat Phys {\bf 5},  110  (2009).

\bibitem{Isenhower10}
L. Isenhower, E. Urban, X.~L. Zhang, A.~T. Gill, T. Henage, T.~A. Johnson,
  T.~G. Walker, and M. Saffman, Phys. Rev. Lett. {\bf 104},  010503  (2010).

\bibitem{Wilk10}
T. Wilk, A. Ga\"etan, C. Evellin, J. Wolters, Y. Miroshnychenko, P. Grangier,
  and A. Browaeys, Phys. Rev. Lett. {\bf 104},  010502  (2010).

\bibitem{Sandoghdar92}
V. Sandoghdar, C.~I. Sukenik, E.~A. Hinds, and S. Haroche, Phys. Rev. Lett.
  {\bf 68},  3432  (1992).

\bibitem{Kuebler10}
H. Kuebler, J.~P. Shaffer, T. Baluktsian, R. Loew, and T. Pfau, Nature
  Photonics {\bf 4},  112   (2010).

\bibitem{Tauschinsky10}
A. Tauschinsky, R.~M.~T. Thijssen, S. Whitlock, H.~B. van Linden van~den
  Heuvell, and R.~J.~C. Spreeuw, Phys. Rev. A {\bf 81},  063411  (2010).

\bibitem{Hinds91}
E.~A. Hinds and V. Sandoghdar, Phys. Rev. A {\bf 43},  398  (1991).

\bibitem{Crosse10}
J.~A. Crosse, S.~A. Ellingsen, K. Clements, S.~Y. Buhmann, and S. Scheel, Phys.
  Rev. A {\bf 82},  010901(R)  (2010).

\bibitem{Saffman05}
M. Saffman and T.~G. Walker, Phys. Rev. A {\bf 72},  022347  (2005).

\bibitem{treutlein04}
P. Treutlein, P. Hommelhoff, T. Steinmetz, T. H\"ansch, and J. Reichel,
  Physical Review Letters {\bf 92},  203005  (2004).

\bibitem{Rekdal04}
P. Rekdal, S. Scheel, P. Knight, and E. Hinds, Physical Review A {\bf {70}},
  013811  (2004).

\bibitem{Hwang11}
J. {Hwang} and E.~A. {Hinds}, ArXiv  1104.3684  (2011).

\bibitem{Kawachi90}
M. Kawachi, Optical and Quantum Electronics {\bf 22},  391  (1990).

\bibitem{Politi08}
A. Politi, M.~J. Cryan, J.~G. Rarity, S. Yu, and J.~L. O'Brien, Science {\bf
  320},  646  (2008).

\bibitem{Long05}
R. Long, T. Rom, W. Hänsel, T.~W. Hänsch, and J. Reichel, The European
  Physical Journal D - Atomic, Molecular, Optical and Plasma Physics {\bf 35},
  125  (2005), 10.1140/epjd/e2005-00177-6.

\bibitem{Petrosyan09}
D. Petrosyan, G. Bensky, G. Kurizki, I. Mazets, J. Majer, and J. Schmiedmayer,
  Phys. Rev. A {\bf 79},  040304  (2009).

\bibitem{Trupke07}
M. Trupke, J. Goldwin, B. Darqui\'{e}, G. Dutier, S. Eriksson, J. Ashmore, and
  E.~A. Hinds, Physical Review Letters {\bf 99},  063601  (2007).

\bibitem{Colombe07}
Y. Colombe, T. Steinmetz, G. Dubois, F. Linke, D. Hunger, and J. Reichel,
  Nature {\bf 450},  272  (2007).

\bibitem{Gleyzes09}
S. Gleyzes, A.~E. Amili, R.~A. Cornelussen, P. Lalanne, C.~I. Westbrook, A.
  Aspect, J. Est{\`e}ve, G. Moreau, A. Martinez, X. Lafosse, L. Ferlazzo, J.~C.
  Harmand, D. Mailly, and A. Ramdane, The European Physical Journal D - Atomic,
  Molecular, Optical and Plasma Physics {\bf 53},  107  (2009).

\end{thebibliography}
\end{document}